\documentclass[final,12pt]{elsarticle}
\usepackage{amsmath}
\usepackage{graphicx}
\usepackage[labelformat=simple]{subcaption}  % 加载subcaption并设置标签格式
\usepackage{amssymb}
\usepackage{amsmath}
\usepackage{graphicx}
\usepackage{subcaption}
\usepackage{booktabs}
\usepackage{xcolor}
\usepackage{ulem}
\usepackage{appendix} % 用于附录格式控制
\usepackage{cleveref}
\newtheorem{proposition}{Proposition}
\journal{XXX}

\begin{document}

\begin{frontmatter}



\title{Exact solutions of the reverse space-time higher-order modified self-steepening nonlinear
Schrödinger equation}


\author[1]{Yanan Wang}
%\ead{wuxihu@ustc.edu.cn}%% Author name
\author[2]{Xi-Hu Wu\corref{cor1}}  
\ead{wuxhu@ustc.edu.cn}% 用 * 标记通讯作者
\cortext[cor1]{Corresponding author}
%\author[2]{.\corref{cor1}}
%\ead{zhangminghe@hrbeu.edu.cn}

\affiliation[1]{organization={School of Mathematical Science},
            addressline={Beihang University}, 
            city={Beijing},
            postcode={102206}, 
            country={China}}
\affiliation[2]{organization={Department of Modern Mechanics},
            addressline={University of Science and Technology of China}, 
            city={Hefei, Anhui},
           postcode={230026}, 
           country={China}}

%\affiliation[2]{organization={School of %Mathematical Science},%Department and Organization 
%addressline={Harbin Engineering University}, 
%city={Harbin},
 %           postcode={150001}, 
  %         country={China}}

%

%% Abstract
\begin{abstract}
This paper investigates a reverse space-time higher-order modified self-steepening nonlinear Schrödinger equation, which distinguishes its standard local counterparts through the reverse space-time symmetry. The integrability of this nonlocal equation is rigorously verified by presenting its associated Lax pair and infinitely many conservation laws. Utilizing the Darboux transformation, we systematically construct a diverse range of localized wave solutions on both zero and nonzero backgrounds. These patterns, such as kinks, exponentially
decaying solitons, asymmetric rogue waves and their interaction solutions, exhibit novel dynamical behaviors that are not found in the local counterparts. This work not only enriches the family of solutions for the equation, but also highlights the effectiveness of the Darboux transformation in exploring nonlinear wave dynamics in nonlocal systems.
\end{abstract}


%\item We develop a new generalized Darboux transformation for the nonlinear coupled dispersionless evolution equation.
%\item We explore degenerate solitons for the nonlinear coupled dispersionless evolution equation and give two novel types of degenerate solitons.
%\item The explicit expressions of the asymptotic solitons for the two types of degenerate solitons are shown.

%% Keywords

\end{frontmatter}

\section{Introduction}
Recently, the parity-time (PT) symmetry of integrable systems has sparked growing interest among researchers in the field of nonlinear science. PT-symmetric systems have been applied in diverse domains, including single-mode lasers, optoelectronic oscillators, sensing technologies and unidirectional transmission systems \cite{bj1,bj2,bj3}. After proposing the integrable reverse-space nonlocal nonlinear Schr\"{o}dinger equation (NLS) \cite{ab1}, Ablowitz and Musslimani investigated some new reverse space–time and reverse time nonlocal nonlinear integrable equations in their subsequent research \cite{ab2}. Subsequently, more nonlocal equations, such as the nonlocal derivative NLS equation \cite{dnls1,dnls2}, the nonlocal modified Korteweg-de Vries equation \cite{mkdv1,mkdv2,mkdv3}, the nonlocal Fokas–Lenells equation \cite{fl1,fl2,fl3}, the nonlocal modified short pulse equation \cite{sp,sp2}, among others, were successively proposed. Meanwhile, several powerful methods, such as Darboux transformation (DT) \cite{dt1,dt2}, inverse scattering transformation \cite{ist1,ist2} and Hirota bilinear method \cite{hi1,hi2}, are applied to solve these nonlocal integrable equations. 

Moreover, PT-symmetric systems, while dissipative in nature, exhibit counter-intuitive conservative properties such as continuous families of nonlinear modes, which distinguish them from traditional systems \cite{pro1}. Additionally, recent studies on nonlocal NLS equations reveal that solitons governed by nonlocal symmetries exhibit fundamentally different dynamics—including recurrent collapse and bounded states over wide parameters—that are not mere superpositions of fundamental solitons \cite{pro2}.

Recently, Wang et al. proposed a new higher-order modified NLS equation with  higher-order dispersion and self-steepening effects \cite{steep}, i.e. 
\begin{align}\label{eq0}
\mathrm{i}q_t&+\mathrm{i}q_{xxx}-3\left(|q|^2q_x+\frac{1}{2} \mathrm{i}|q|^4q\right)_x+\left(2q_{xx}-3|q|^4q\right)\rho+\left(2\mathrm{i}q^2q^*_x+10\mathrm{i}|q|^2q_x\right)\rho \notag\\&+\left(4|q|^2q+4\mathrm{i}q_x-8q\right)\rho^2 +8q\rho^3=0,
\end{align}
where $q$ is a complex-valued function and the subscripts denote the 
corresponding partial derivatives, the asterisk denotes the complex conjugate and $\rho$ is a real parameter. Eq.\eqref{eq0} can be derived via the specific reduction condition $r=-q^*$ from an integrable coupled system given in Appendix \eqref{eq:app1}. 
Thereby, applying the symmetry reduction $r(x,t)=-q(-x,-t)$, which is nonlocal both in
space and time, we can introduce the following new integrable reverse space-time nonlocal equation as follows,
\begin{align}\label{eq}
\mathrm{i}q_t&+\mathrm{i}q_{xxx}-3\left[qq(-x,-t)q_x+\frac{1}{2} \mathrm{i}q^3q^2(-x,-t)\right]_x+\left[2q_{xx}-3q^3q^2(-x,-t)\right]\rho\notag\\&+\left[-2\mathrm{i}q^2q_x(-x,-t)+10\mathrm{i}qq(-x,-t)q_x\right]\rho+\left[4q^2q(-x,-t)+4\mathrm{i}q_x-8q\right]\rho^2 \notag\\&+8q\rho^3=0,
\end{align}
where $q_x(-x,-t)$ denotes one first differentiate with respect to $x$ and then replace $x\to -x,t\to-t$.
The reverse space-time modified NLS equation has extensive physical applications in diverse fields such as optics, ocean waves, quantum entanglement, and magnetic systems \cite{app1,app2}. The nonlocal equation can not only extend the solutions of local equations to a more general situation but also advance the physical understanding the formation mechanism of rogue wave formation mechanisms.

In our work, we employ the DT method to explore multiple localized wave solutions of Eq.\eqref{eq} on zero and nonzero backgrounds. This paper is organized as follows. In Section \ref{1s}, a Lax pair and infinitely many conservation laws are given, which further confirm the integrability of Eq.\eqref{eq}. In Section \ref{2s}, a DT for Eq.\eqref{eq} is constructed. In Section \ref{3s}, the soliton, breather, rogue wave solutions and related interaction solutions are shown through degenerate DT and the semi-degenerate DT. The corresponding dynamical characteristics and evolutionary behaviors are discussed. In Section \ref{4s}, the conclusions and discussions are given.

\section{Lax pair and conservation laws}\label{1s} 
It's well-known that Eq.\eqref{eq0} is derived from a coupled system given in Appendix \eqref{eq:app1} using the reduction condition $r=-q^*$. Appendix \eqref{eq:app2} provides the Lax pair for the coupled system. Motivated by \cite{gi,lpd}, the following Lax pair for Eq.\eqref{eq} is obtained via the nonlocal reduction $r(x,t)=-q(-x,-t)$.
\begin{align}\label{lax}
\Phi_x=
U\Phi, ~~~\Phi_t=V
\Phi,
\end{align}
where 
\begin{align*}
U&=\begin{pmatrix}
-\dfrac{i}{\lambda^2} + \rho i & \dfrac{q}{\lambda} \\
-\dfrac{q(-x,-t)}{\lambda} & \dfrac{i}{\lambda^2} - \rho i
\end{pmatrix}, ~~V=\begin{pmatrix}
V_1 & V_2 \\
V_3 & -V_1
\end{pmatrix},
\end{align*}
\begin{align*}
V_1=&-\frac{3i q^2 q^2(-x,-t)}{2}\lambda^{-2} - 4i \rho^2 -q_x(-x,-t) \, q\lambda^{-2} - q_x \, q(-x,-t)\lambda^{-2} \\&+ 2i q q(-x,-t)\lambda^{-4} + 8i \rho\lambda^{-4} - 4i\lambda^{-6}, \\
V_2=&-q_{xx}\lambda^{-1} - 3i q_x q q(-x,-t)\lambda^{-1} + 2i q_x\lambda^{-3} + \frac{3 q^3 q^2(-x,-t)}{2}\lambda^{-1}- 4 q \rho^2\lambda^{-1} \\&-2 q^2 q(-x,-t) \rho\lambda^{-1} - 2 q^2 q(-x,-t)\lambda^{-3} - 4 q \rho\lambda^{-3} + 4 q\lambda^{-5}, \\
V_3=&q_{xx}(-x,-t)\lambda^{-1} -3iq_x(-x,-t) qq(-x,-t)\lambda^{-1} - 2i q_x(-x,-t)\lambda^{-3}\\
&- \frac{3 q^2 q^3(-x,-t)}{2}\lambda^{-1} + 2 q q^2(-x,-t) \rho\lambda^{-1} + 2 q q^2(-x,-t)\lambda^{-3}\\& + 4 q(-x,-t) \rho^2\lambda^{-1} + 4 q(-x,-t) \rho\lambda^{-3} - 4 q(-x,-t)\lambda^{-5},
\end{align*}
where $\Phi=(\phi_1,\phi_2)^{T}$ is the vector eigenfunction.
It is rigorously demonstrated that Eq.\eqref{eq} follows from the zero curvature equation $U_t-V_x+[U,V]=0$. 

Subsequently, starting with the Lax pair Eq.\eqref{lax} and according to \cite{sh1,sh2}, we present the corresponding infinitely many conservation laws. Introducing the complex function $\Lambda=\frac{\phi_2}{\phi_1}$, we obtain the following Riccati-type equation,
\begin{align}\label{Ri}
    \Lambda_x=-q(-x,-t)\lambda^{-1}-(2i\rho-2i\lambda^{-2})\Lambda-q\lambda\Lambda^2.
\end{align}
From Eq.\eqref{Ri}, it is clear that $\phi_1$ and $\phi_2$ depend on the parameters $\lambda$, $\rho$ and the solution $q$. We assume the expansion $\Lambda=\sum_{k=1}^{\infty}\Lambda_{2k-1}\lambda^{2k-1}$, where $\Lambda_{2k-1}$ is a function dependent on both $x$ and $t$ to be determined. Then substituting it into Eq.\eqref{Ri} and equating the coefficients corresponding to identical powers of $\lambda$ to zero yields the following results:
\begin{align*}
    \lambda^{-1}:&~~\Lambda_1=-\frac{i}{2}q(-x,-t),\\
    \lambda:&~~\Lambda_3=-\frac{i}{2}(2i\rho\Lambda_1+q\Lambda_1^2+\Lambda_{1,x})
    \\ &\quad\quad=\frac{iqq^2(-x,-t)}{8}-\frac{i\rho q(-x,-t)}{2}+\frac{q_x(-x,-t)}{4},\\
    \lambda^{3}:&~~\Lambda_5=-\frac{i}{2}(2i\rho\Lambda_3+2q\Lambda_1\Lambda_3+\Lambda_{3,x})
    \\&\quad\quad=-\frac{i}{16}q^2q^3(-x,-t)-\frac{1}{4}qq(-x,-t)q_x(-x,-t)+\frac{1}{16}q_xq^2(-x,-t)\\&\quad\quad\quad+\frac{3i\rho}{8}qq^2(-x,-t)+\frac{\rho}{2} q_x(-x,-t)-\frac{i\rho^2}{2}q(-x,-t)+\frac{i}{8}q_{xx}(-x,-t), 
    \end{align*}
\begin{align*}
    \lambda^{5}:&~~\Lambda_7=-\frac{i}{2}(2i\rho\Lambda_5+2q\Lambda_1\Lambda_5+q\Lambda_3^2+\Lambda_{5,x}),
    \\& \vdots\\
    \lambda^{2k-3}:&~~\Lambda_{2k-1}=\left\{
\begin{aligned}
&-\frac{i}{2}(2i\rho\Lambda_{2k-3}+2q\sum_{j=1}^{k/2-1}\Lambda_{2j-1}\Lambda_{2k-2j-1}+q\Lambda_{k-1}^2+\Lambda_{2k-3,x}),~~ \text{k is even}, \\
&-\frac{i}{2}(2i\rho\Lambda_{2k-3}+2q\sum_{j=1}^{(k-1)/2}\Lambda_{2j-1}\Lambda_{2k-2j-1}+\Lambda_{2k-3,x}),  ~~ \text{k is odd}.
\end{aligned}
\right.
\end{align*}
Based on the compatibility condition $(\ln\phi)_{xt}=(\ln\phi)_{tx}$,
infinitely many conservation laws for Eq.\eqref{eq} are expressed in the form
\begin{align}
   \frac{\partial D_k}{\partial t}= \frac{\partial F_k}{\partial x}
\end{align}
where $D_k$ and $F_k$ are the conserved densities and fluxes, respectively, with the explicit expressions,
\begin{align*}
    D_1&=q\Lambda_1,\\
    F_1&=\frac{1}{2}q[3q^2q^2(-x,-t)-4\rho qq(-x,-t)-8\rho^2]\Lambda_1+\frac{1}{2}[-4qq(-x,-t)-8\rho]\Lambda_3+4q\Lambda_5,\\
    D_2&=q\Lambda_3,\\
    F_2&=\frac{1}{2}q[3q^2q^2(-x,-t)-4\rho qq(-x,-t)-8\rho^2]\Lambda_3+\frac{1}{2}[-4qq(-x,-t)-8\rho]\Lambda_5+4q\Lambda_7,\\
    \vdots\\
    D_k&=q\Lambda_{2k-1},\\
    F_k&=\frac{1}{2}q[3q^2q^2(-x,-t)-4\rho qq(-x,-t)-8\rho^2]\Lambda_{2k-1}+\frac{1}{2}[-4qq(-x,-t)-8\rho]\Lambda_{2k+1}+4q\Lambda_{2k+3},\\
    \vdots
\end{align*}

In conclusion, the systematic derivation of the Lax pair and an infinite number of conservation laws further confirm the integrability of Eq.\eqref{eq}, providing theoretical support for the subsequent solution construction.

\section{Darboux transformation}\label{2s}
In this section, we construct a DT for the reverse space-time nonlocal mNLS equation \eqref{eq}. We begin with the Lax pair \eqref{eq:app2} of the coupled system and introduce a gauge transformation $\Phi[1]=T_1\Phi$. Under this transformation, the new Lax pair for the spectral function $\Phi[1]$ takes the form 
\begin{align*}
    \Phi[1]_x&=(T_{1,x}+T_1U)T_1^{-1}\Phi[1]=U[1]\Phi[1],\\
    \Phi[1]_t&=(T_{1,t}+T_1V)T_1^{-1}\Phi[1]=V[1]\Phi[1],
\end{align*}
where $U[1]$ and $V[1]$ are derived by substituting $q,r$ with $q[1],r[1]$ in the original spectral matrices $U$ and $V$. Following \cite{steep}, we assume the Darboux matrix to be
\begin{align*}
    T(\lambda)=\begin{pmatrix}
        \lambda^{2N}+\sum_{j=0}^{N-1}A_{1,j}\lambda^{2j} & \sum_{j=0}^{N-1}A_{2,j}\lambda^{2j+1} \\
        \sum_{j=0}^{N-1}A_{3,j}\lambda^{2j+1} & \lambda^{2N}+\sum_{j=0}^{N-1}A_{4,j}\lambda^{2j}
    \end{pmatrix},
\end{align*}
where $A_{i,j}, (i=1,2,3,4)$ are functions of $x$ and $t$ to be determined. Subsequently, using the fact $T_1\Phi_1=0$, where $\Phi_1=(\phi_{11},\phi_{21})^T$ is the eigenfunction of the Lax pair \eqref{eq:app2} under the spectral parameter $\lambda=\lambda_1$, we derive the following proposition.
\begin{proposition}\label{pro1}
    Assume that $\Phi_{k}=(\phi_{1,k},\phi_{2,k})^T$ are the eigenfunctions of \eqref{eq:app2} corresponding to the spectral parameters $\lambda=\lambda_i$. Then the $N$-fold DT of the coupled system is given by
    \begin{align*}
        q[N]&=q-2i\rho A_{2,N-1}+A_{2,N-1,x},\\
        r[N]&=r+2i\rho A_{3,N-1}+A_{3,N-1,x},
    \end{align*}
    where 
    $A_{2,N-1}=\frac{|B_{11}|}{|B_{12}|},A_{3,N-1}=\frac{|B_{21}|}{|B_{22}|}$,
    with 
  \[ B_{12}=
    \begin{pmatrix}
        \lambda_1^{2N-2}\phi_{1,1} &  \lambda_1^{2N-1}\phi_{2,1} &  \lambda_1^{2N-4}\phi_{1,1} &
        \lambda_1^{2N-3}\phi_{2,1} & \cdots & \phi_{1,1} & \lambda_1\phi_{2,1} \\
         \lambda_2^{2N-2}\phi_{1,2} &  \lambda_2^{2N-1}\phi_{2,2} &  \lambda_2^{2N-4}\phi_{1,2} &
        \lambda_2^{2N-3}\phi_{2,2} & \cdots & \phi_{1,2} & \lambda_2\phi_{2,2}  \\
        \vdots & \vdots & \vdots & \vdots & \ddots & \vdots & \vdots \\
        \lambda_{2N}^{2N-2}\phi_{1,2N} &  \lambda_{2N}^{2N-1}\phi_{2,2N} &  \lambda_{2N}^{2N-4}\phi_{1,2N} &
        \lambda_{2N}^{2N-3}\phi_{2,2N} & \cdots & \phi_{1,2N} & \lambda_{2N}\phi_{2,2N} 
    \end{pmatrix},\]
    \[B22= \begin{pmatrix}
     \lambda_1^{2N-1}\phi_{1,1} & \lambda_1^{2N-2}\phi_{2,1} & \lambda_1^{2N-3}\phi_{1,1} & \cdots & \lambda_1\phi_{1,1} & \phi_{2,1} \\
     \lambda_2^{2N-1}\phi_{1,2} & \lambda_2^{2N-2}\phi_{2,2} & \lambda_2^{2N-3}\phi_{1,2} & \cdots & \lambda_2\phi_{1,2} & \phi_{2,2} \\
     \vdots & \vdots & \vdots & \ddots & \vdots & \vdots \\
     \lambda_{2N}^{2N-1}\phi_{1,2N} & \lambda_{2N}^{2N-2}\phi_{2,2N} & \lambda_{2N}^{2N-3}\phi_{1,2N} & \cdots & \lambda_{2N}\phi_{1,2N} & \phi_{2,2N} 
    \end{pmatrix},
    \]
 $B_{11}$ is obtained by replacing the second column of $B_{12}$ with 
$$(-\lambda_1^{2N}\phi_{1,1}, -\lambda_2^{2N}\phi_{1,2}, - \lambda_3^{2N}\phi_{1,3},\cdots,-\lambda_{2N}^{2N}\phi_{1,2N} )^{T},$$
and $B_{21}$ is derived by replacing the first column of $B_{22}$ with 
$$(-\lambda_1^{2N}\phi_{2,1}, -\lambda_2^{2N}\phi_{2,2}, - \lambda_3^{2N}\phi_{2,3},\cdots,-\lambda_{2N}^{2N}\phi_{2,2N} )^T.$$
\end{proposition}

Building on the above result, we now add the nonlocal symmetry reduction $r(x,t)=-q(-x,-t)$. This means that the eigenfunctions need to meet the property: $\phi_{1,k}(x,t)=\phi_{2,k}(-x,-t)$. The DT for the reverse space-time nonlocal equation \eqref{eq} is thereby established.

\section{Exact solutions on different backgrounds}\label{3s}
In this section, various exact localized wave solutions are exhibited from zero and nonzero seed solutions, respectively. These structures are richer than those found in local equations.
\subsection{zero background}
When choosing the seed solution $r(x,t)=-q(-x,-t)=0$, we can derive the following eigenfunctions, 
\begin{align*}
    \Phi_{k}=\begin{pmatrix}
        \phi_{1,k} \\ \phi_{2,k}
    \end{pmatrix}=
    \begin{pmatrix}
        e^{i[(\lambda_k^2\rho-1)\lambda_k^{-2}x-4(\rho^2\lambda_k^6-2\lambda_k^2\rho+1)\lambda_k^{-6}t]}\\ 
        e^{-i[(\lambda_k^2\rho-1)\lambda_k^{-2}x-4(\rho^2\lambda_k^6-2\lambda_k^2\rho+1)\lambda_k^{-6}t]}
    \end{pmatrix}.
\end{align*}
It can be verified that this eigenfunction satisfies condition $\phi_{1,k}(x,t)=\phi_{2,k}(-x,-t)$. Thus, when $N=1$, the exact expression of the solution is 
\begin{align}\label{1-so}
    q[1]=-\frac{2i(\lambda_1^2-\lambda_2^2)\left[\lambda_1e^{\theta_1(\lambda_1)}-\lambda_2e^{\theta_1(\lambda_2)}\right]}{\lambda_1\lambda_2(\lambda_2e^{-\theta_2}-\lambda_1e^{\theta_2})^2},
\end{align}
where $\theta_1(\lambda)=-2i[(4\lambda^6\rho^2-8\lambda^2\rho+4)t-(\lambda^6\rho-\lambda^4)x]\lambda^{-6}$ and $\theta_2=i[((8\lambda_1^2\rho-4)\lambda_2^4+(8\lambda_1^2\rho-4)\lambda_1^2\lambda_2^2-4i\lambda_1^4)t-\lambda_1^4\lambda_2^4x](\lambda_1^2-\lambda_2^2)\lambda_1^6\lambda_2^6$.
We find that the parameter $\rho$, associated with coefficients of $x$ and $t$, exists in the phase part of the exponential term. This implies that $\rho$ may exert influences on the dynamical behaviors of wave solutions from different aspects, such as the background field, nonlinearity strength, dispersion relation, phase and dissipative effects. 

\begin{figure}[ht!]
    \centering
   \begin{subfigure}{0.35\textwidth}
        \centering
        \includegraphics[width=\textwidth]{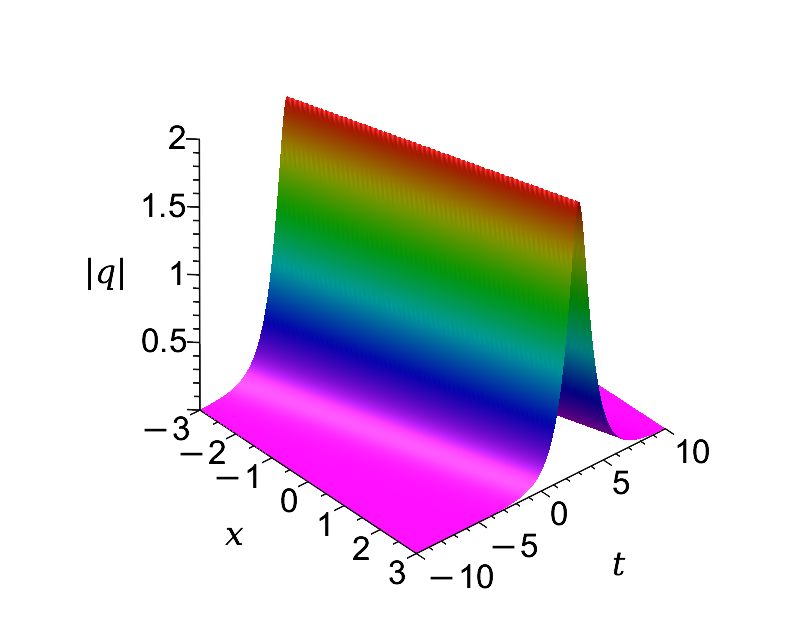}
        \caption{}
        \label{fg11}
    \end{subfigure}
  %  \hspace{1cm}
    \begin{subfigure}{0.35\textwidth}
        \centering
        \includegraphics[width=\textwidth]{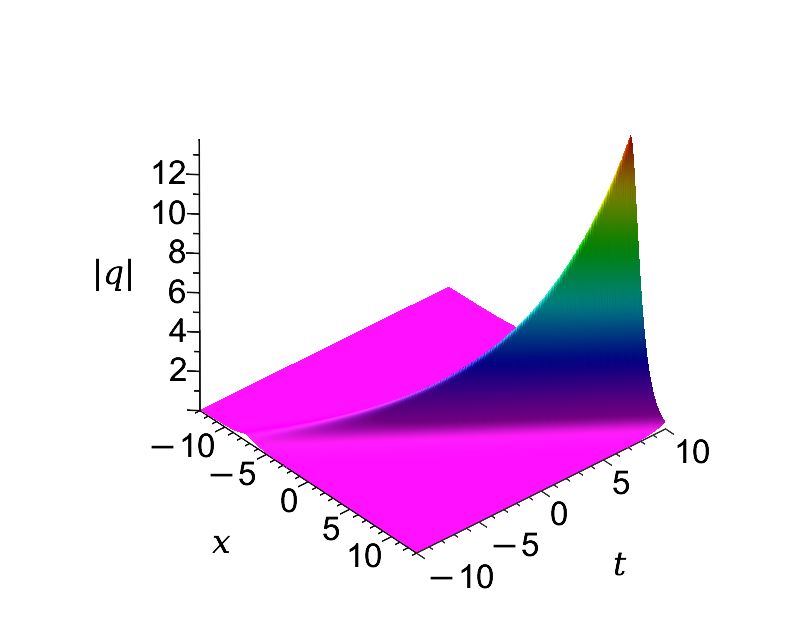}
        \caption{}
        \label{fg12}
    \end{subfigure}
    \begin{subfigure}{0.26\textwidth}
        \centering
        \includegraphics[width=\textwidth]{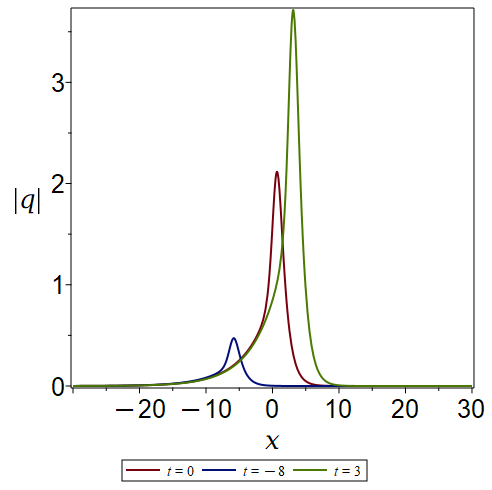}
        \caption{}
        \label{fg13}
    \end{subfigure}
     \caption{(a) The bell-shaped soliton with $\rho=0,\lambda_1=\lambda_2^*=1+\mathrm{i}$; (b) The exponentially decaying soliton with $\rho=0,\lambda_1=1+\mathrm{i}, \lambda_2=2-2\mathrm{i}$; (c) The cross-sectional view of (b) at $t=-8, t=0, t=3$.}
    \label{fg1}
\end{figure}

Based on the solution \eqref{1-so}, two types of solitons are obtained if $\mathrm{Re}(\lambda_k)\neq0,\mathrm{Im}(\lambda_k)\neq0,k=1,2$. Then when $\lambda_1=\lambda_2^*$ holds, a bell-shaped soliton with strict spatial localization and dynamical stability is shown in Fig.\ref{fg11}. There exists the soliton whose amplitude decays exponentially over time when $\lambda_1\neq\lambda_2^*$ holds as is exhibited in Fig.\ref{fg12} and Fig.\ref{fg13}.

\begin{figure}[htbp]
    \centering
   \begin{subfigure}{0.35\textwidth}
        \centering
        \includegraphics[width=\textwidth]{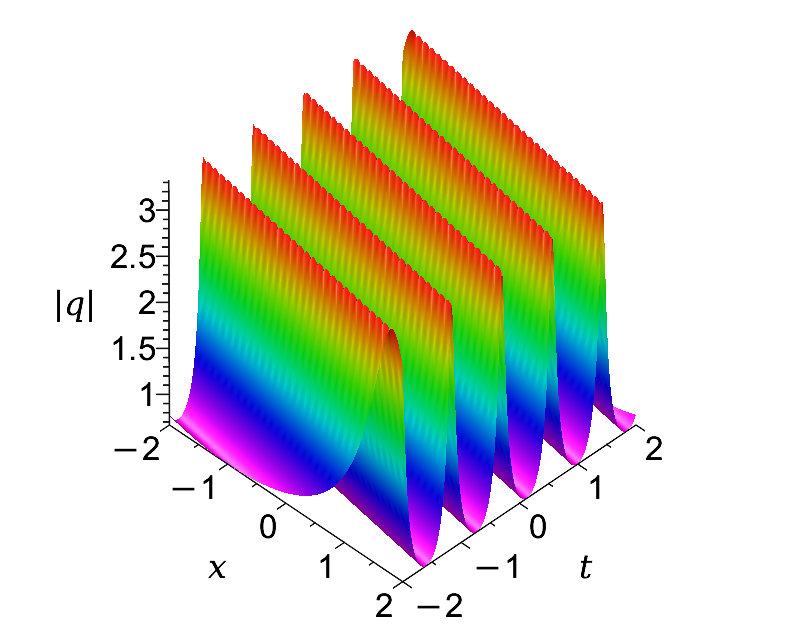}
        \caption{}
        \label{fg21}
    \end{subfigure}
  %  \hspace{1cm}
    \begin{subfigure}{0.35\textwidth}
        \centering
        \includegraphics[width=\textwidth]{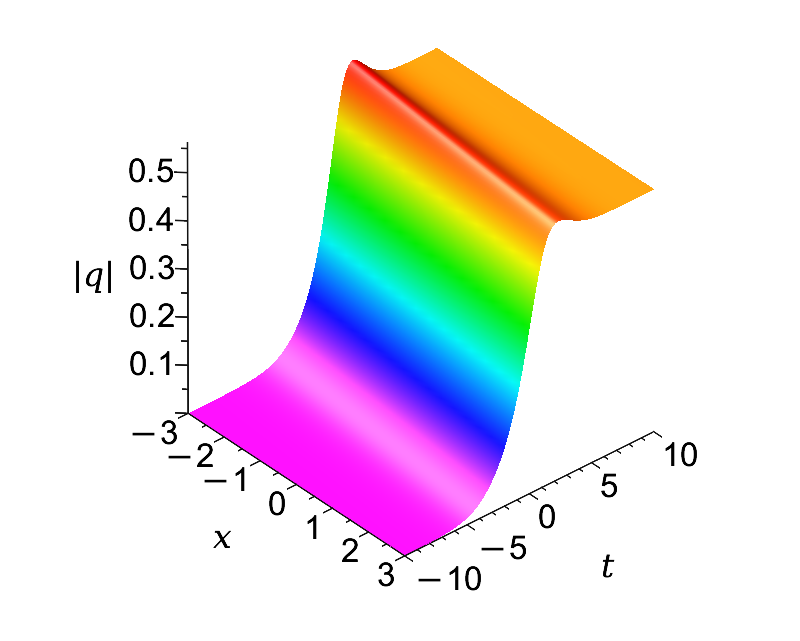}
        \caption{}
        \label{fg22}
    \end{subfigure}
    \begin{subfigure}{0.26\textwidth}
        \centering
        \includegraphics[width=\textwidth]{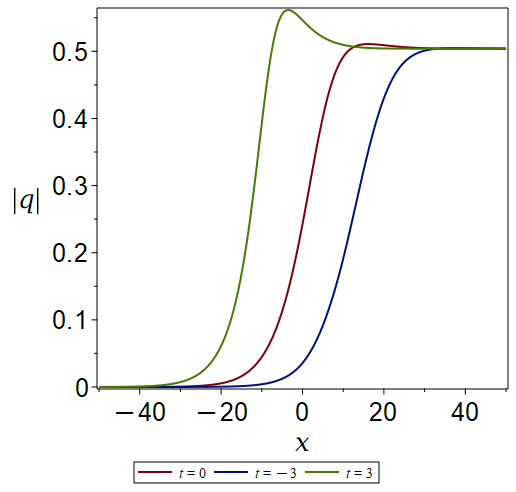}
        \caption{}
        \label{fg23}
    \end{subfigure}
     \caption{(a) The periodic wave solution with $\rho=0, \lambda_1=\frac{3}{2}, \lambda_2=1$; (b) The kink solution with $\rho=1, \lambda_1=2\mathrm{i}, \lambda_2=\frac{1}{2}-2\mathrm{i}$; (c) The cross-sectional view of (b) at $t=-3, t=0, t=3$.}
    \label{fg2}
\end{figure}

Periodic solutions are obtained when $\mathrm{Im}(\lambda_1)=\mathrm{Im}(\lambda_2)=0$ or $\mathrm{Re}(\lambda_1)=\mathrm{Re}(\lambda_2)=0$, as depicted in Fig.\ref{fg21}. Furthermore, when setting $\mathrm{Re}(\lambda_1)=0,\mathrm{Re}(\lambda_2)\neq0$, Fig.\ref{fg22} and \ref{fg23} present a kink solution, which does not exist in the local equation \eqref{eq0}. We can find that the waveform propagates rightward over time, with its prominent peak gradually diminishing and eventually evolving towards a stable constant amplitude.

\begin{figure}[ht!]
    \centering
   \begin{subfigure}{0.3\textwidth}
        \centering
        \includegraphics[width=\textwidth]{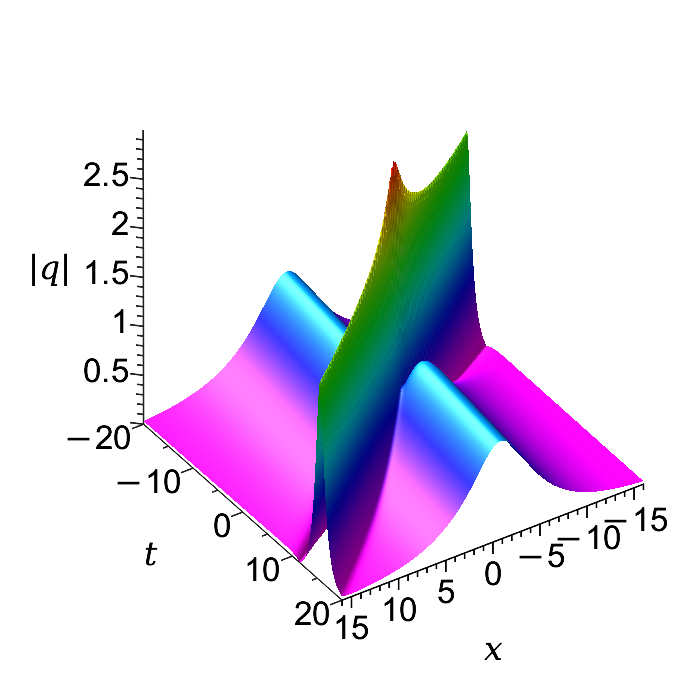}
        \caption{}
        \label{fg31}
    \end{subfigure}
  %  \hspace{1cm}
    \begin{subfigure}{0.3\textwidth}
        \centering
        \includegraphics[width=\textwidth]{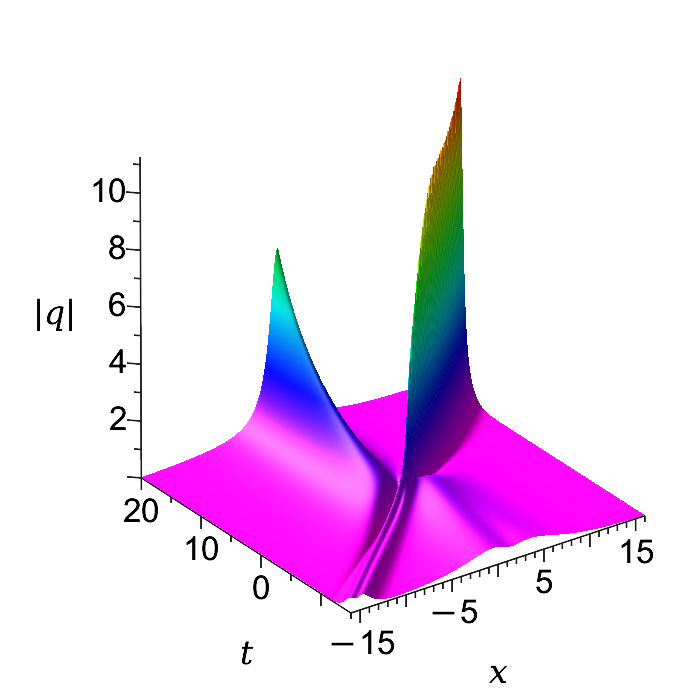}
        \caption{}
        \label{fg32}
    \end{subfigure}
    \begin{subfigure}{0.35\textwidth}
        \centering
        \includegraphics[width=\textwidth]{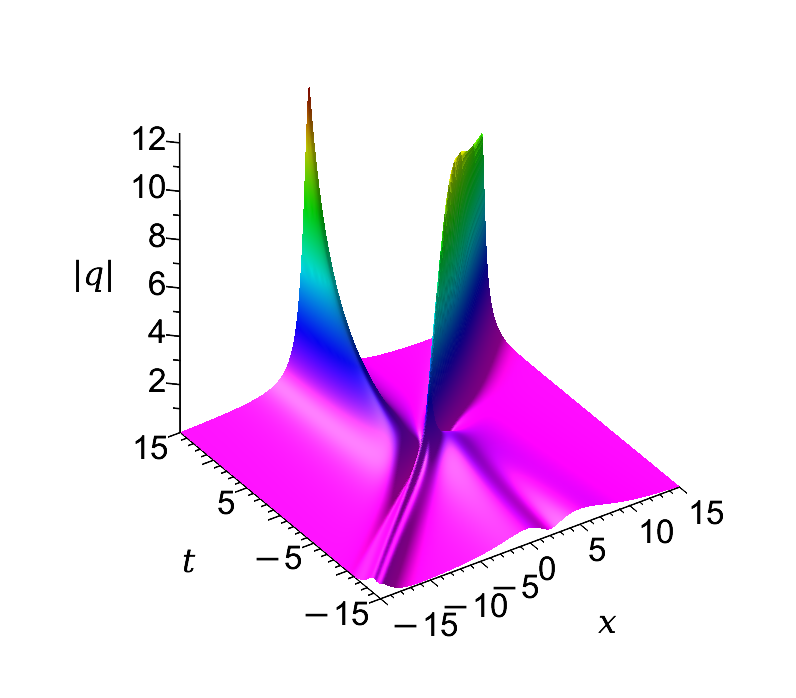}
        \caption{}
        \label{fg33}
    \end{subfigure}
     \caption{The interaction solutions with $\rho=0, \lambda_1=1+\mathrm{i}, \lambda_2=2-2\mathrm{i} $. (a) Two classical solitons with $\lambda_3=1-\mathrm{i}, \lambda_4=2+2\mathrm{i}$; (b) The classical soliton and the exponentially decaying soliton with $\lambda_3=1-\mathrm{i}, \lambda_4=2+\mathrm{i}$; (c) Two exponentially decaying solitons with $\lambda_3=1-1.01\mathrm{i}, \lambda_4=2+1.25\mathrm{i}$.}
    \label{fg3}
\end{figure}

Finally, we consider the interaction solutions under zero background for the case $N=2$. The second-order solution involves four spectral parameters. Thus, when $\mathrm{Re}(\lambda_k)\neq0, \mathrm{Im}(\lambda_k)\neq0, k=1,2,3,4$, we explore the interaction solutions between solitons. Fig.\ref{fg31} shows the interaction between two classical solitons, and Fig.\ref{fg32} shows the interaction between two exponentially decaying solitons. Naturally, the interaction between the classical soliton and the exponentially decaying soliton is exhibited in Fig.\ref{fg33}.

\begin{figure}[ht!]
    \centering
   \begin{subfigure}{0.31\textwidth}
        \centering
        \includegraphics[width=\textwidth]{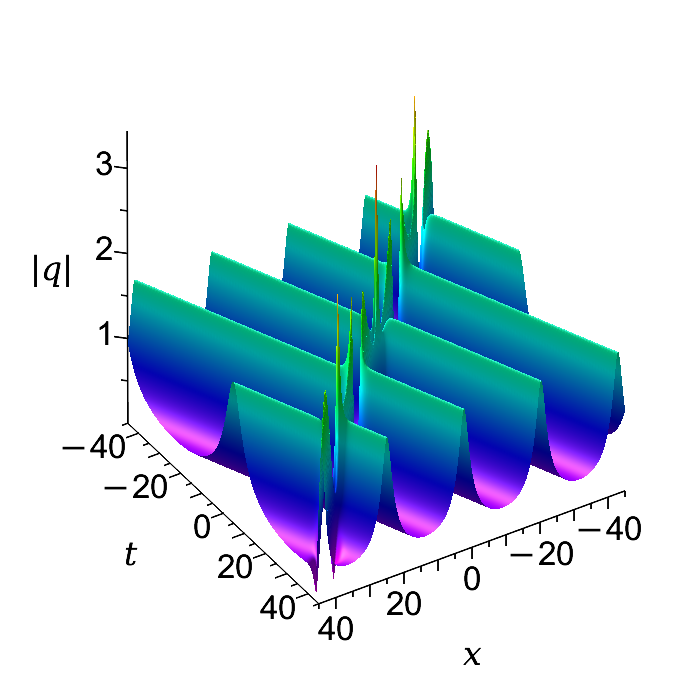}
        \caption{}
        \label{fg41}
    \end{subfigure}
  %  \hspace{1cm}
    \begin{subfigure}{0.31\textwidth}
        \centering
        \includegraphics[width=\textwidth]{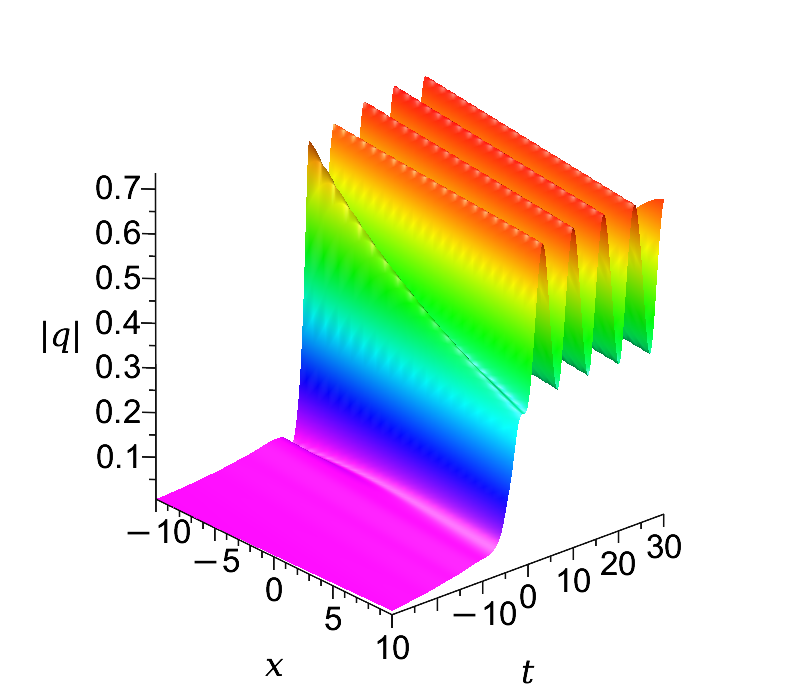}
        \caption{}
        \label{fg42}
    \end{subfigure}
    \begin{subfigure}{0.34\textwidth}
        \centering
        \includegraphics[width=\textwidth]{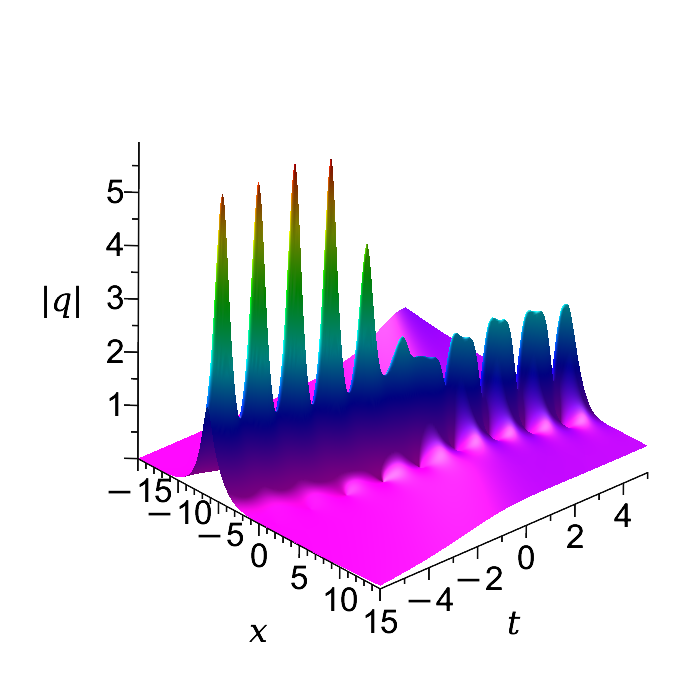}
        \caption{}
        \label{fg43}
    \end{subfigure}
     \caption{The interaction solutions. (a) The soliton and the periodic wave with $\rho=0, \lambda_1=\lambda_2^*=1+\mathrm{i}, \lambda_3=2, \lambda_4=3$; (b) The kink and the periodic wave with $\rho=1, \lambda_1=2\mathrm{i}, \lambda_2=3\mathrm{i}, \lambda_3=\frac{1}{2}-2\mathrm{i}, \lambda_4=\frac{1}{3}-3\mathrm{i}$; (c) The kink and the breather-like solution with $\rho=1, \lambda_1=2\mathrm{i},\lambda_2=\frac{1}{2}-2\mathrm{i}, \lambda_3=\lambda_4^*=1-\mathrm{i}$.}
    \label{fg4}
\end{figure}

When the spectral parameters take other values, more diverse interaction solutions are obtained. As presented in Fig.\ref{fg4}, these include interactions between a periodic wave and a soliton, a kink and a periodic wave, and a kink and a breather-like wave. Here, the breather-like solution refers to a soliton characterized by periodic oscillatory behavior, reflecting typical breathing dynamics.

\subsection{nonzero background}
\begin{figure}[htbp]
    \centering
   \begin{subfigure}{0.3\textwidth}
        \centering
        \includegraphics[width=\textwidth]{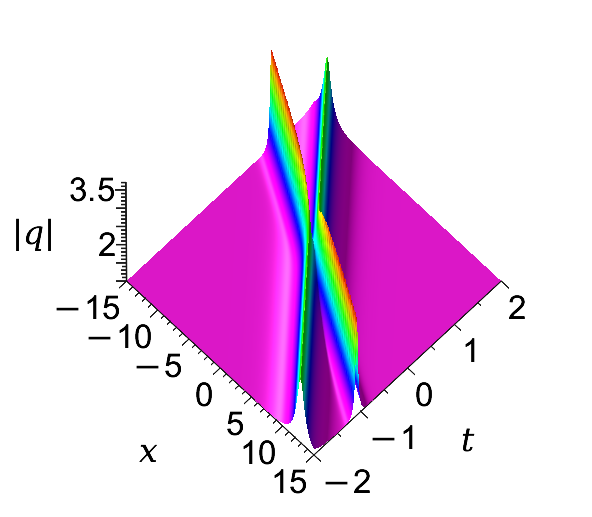}
        \caption{}
        \label{fg51}
    \end{subfigure}
   \hspace{0.1cm}
    \begin{subfigure}{0.3\textwidth}
        \centering
        \includegraphics[width=\textwidth]{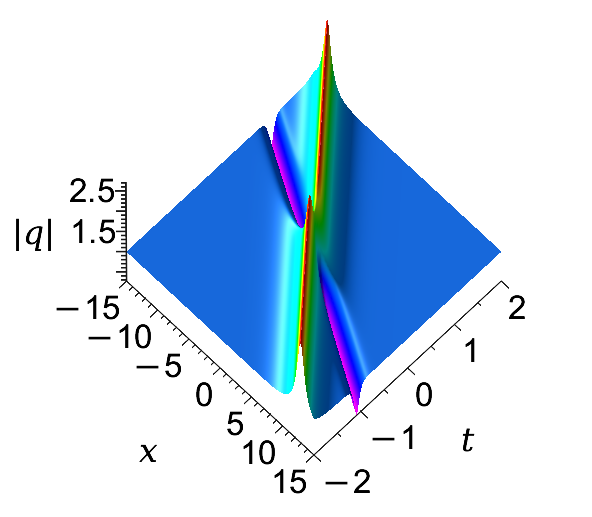}
        \caption{}
        \label{fg52}
    \end{subfigure}
     \hspace{0.1cm}
    \begin{subfigure}{0.3\textwidth}
        \centering
        \includegraphics[width=\textwidth]{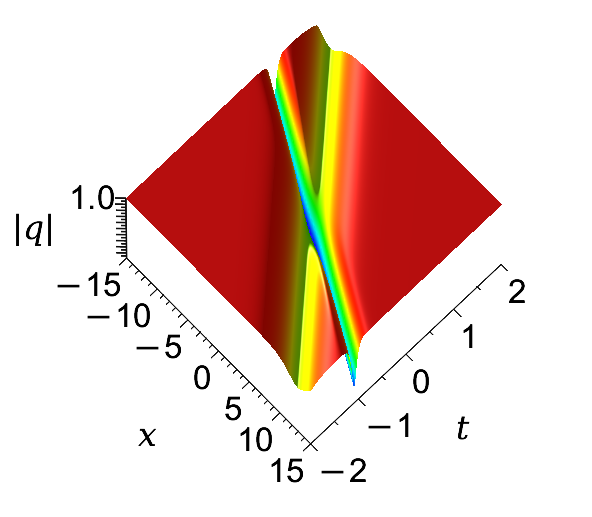}
        \caption{}
        \label{fg53}
    \end{subfigure}
     \caption{(a) The interaction solutions with $\rho=0, a=1, b=\frac{11}{2}, c=1, c_1=c_2=1$. (a) Two bright solitons with $\lambda_1=-\mathrm{i}, \lambda_2=2\mathrm{i}$; (b) The bright soliton and the dark soliton with $\lambda_1=\mathrm{i}, \lambda_2=2\mathrm{i}$; (c) Two dark solitons with $\lambda_1=\mathrm{i}, \lambda_2=-2\mathrm{i}$.}
    \label{fg5}
\end{figure}

In this section, we mainly explore the first-order solutions of the reverse space-time nonlocal equation \eqref{eq} under nonzero background. We begin with the nonzero seed solution $q(x,t)=-r(-x,-t)=ce^{-i(ax+bt)}$ and substitute them into the Lax pair \eqref{eq:app2} to obtain the following eigenfunctions,
\begin{align*}
    \Phi_{1,k}&=\begin{pmatrix}
        \psi_{1,k} \\ \psi_{2,k} 
    \end{pmatrix}=
    \begin{pmatrix}
        c_1e^{-(2\lambda_k^4x-At)\lambda_k^{-6}H_k+\frac{1}{2}i(ax+bt)} \\
        c_2\frac{2i+i(a-2\rho)\lambda_k^2-H_k}{2\lambda_k c}e^{-(2\lambda_k^4x-At)\lambda_k^{-6}H_k-\frac{1}{2}i(ax+bt)}
    \end{pmatrix}, \\
    \Phi_{2,k}&=\begin{pmatrix}
        \psi_{3,k} \\ \psi_{4,k} 
    \end{pmatrix}=
    \begin{pmatrix}
        c_1e^{(2\lambda_k^4x-At)\lambda_k^{-6}H_k+\frac{1}{2}i(ax+bt)} \\
        c_2\frac{2i+i(a-2\rho)\lambda_k^2+H_k}{2\lambda_k c}e^{(2\lambda_k^4x-At)\lambda_k^{-6}H_k-\frac{1}{2}i(ax+bt)}
    \end{pmatrix},
\end{align*}
where $c_1,c_2$ are the real coefficients and $A=\frac{1}{4}[(-3c^4-6ac^2+4c^2\rho-2a^2+8\rho^2)\lambda_k^4+4(c^2+a+2\rho)\lambda_k^2-8]t-2\lambda_k^4x$, $H_k=\sqrt{-4-(a-2\rho)^2\lambda_k^4+(-4c^2-4a+8\rho)\lambda_k^2}$.
To ensure the validity of the nonlocal reduction condition $r(x,t)=-q(-x,-t)$, a new eigenfunction is constructed using the linear superposition principle,
\begin{align}\label{eg}
    \Phi_k=\begin{pmatrix}
        \phi_{1,k} \\ \phi_{2,k}
    \end{pmatrix}=\begin{pmatrix}
        \psi_{1,k}+\psi_{3,k}+\psi_{2,k}(-x,-t)+\psi_{4,k}(-x,-t) \\
        \psi_{2,k}+\psi_{4,k}+\psi_{1,k}(-x,-t)+\psi_{3,k}(-x,-t) 
    \end{pmatrix}.
\end{align}
According to Proposition \ref{pro1}, when $\mathrm{Re}(\lambda_1)=\mathrm{Re}(\lambda_2)=0$, different types of localized wave solutions are derived. Firstly, bright-bright, bright-dark and dark-dark soliton solutions are obtained as shown in Fig.\ref{fg5}. It is worth noting that they are constructed via a single-fold DT, unlike the interaction solutions generated by double-fold DT.

\begin{figure}[ht!]
    \centering
   \begin{subfigure}{0.33\textwidth}
        \centering
        \includegraphics[width=\textwidth]{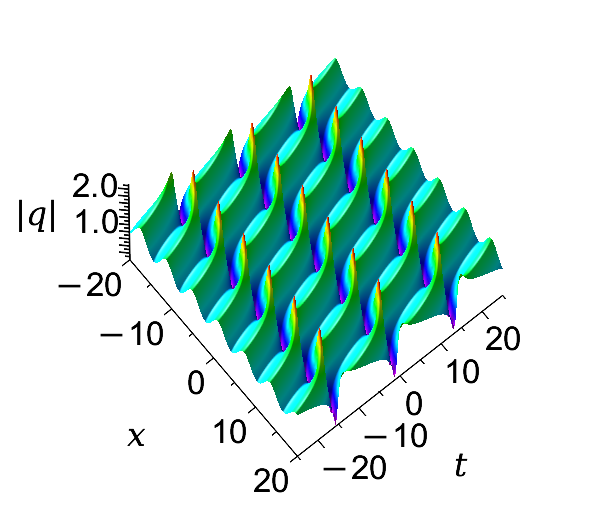}
        \caption{}
        \label{fg61}
    \end{subfigure}
  %  \hspace{1cm}
    \begin{subfigure}{0.32\textwidth}
        \centering
        \includegraphics[width=\textwidth]{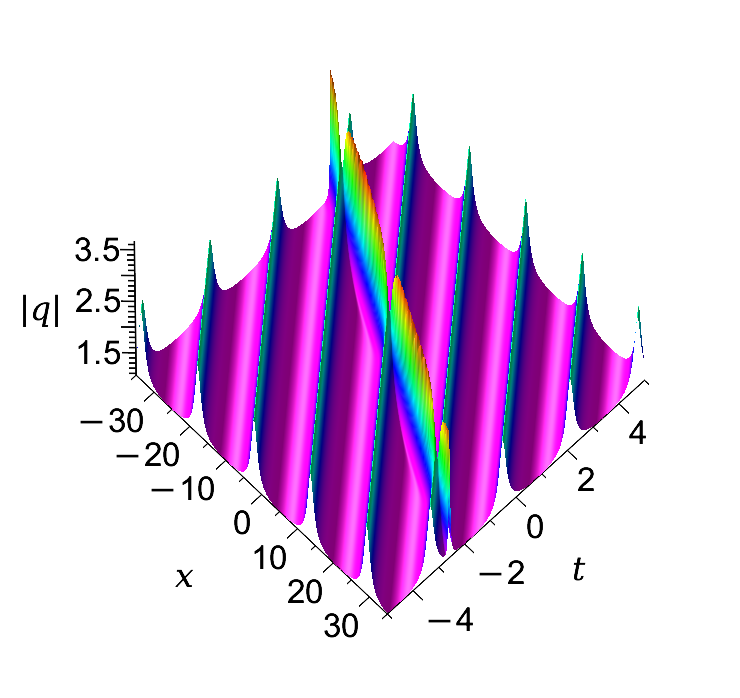}
        \caption{}
        \label{fg62}
    \end{subfigure}
    \begin{subfigure}{0.32\textwidth}
        \centering
        \includegraphics[width=\textwidth]{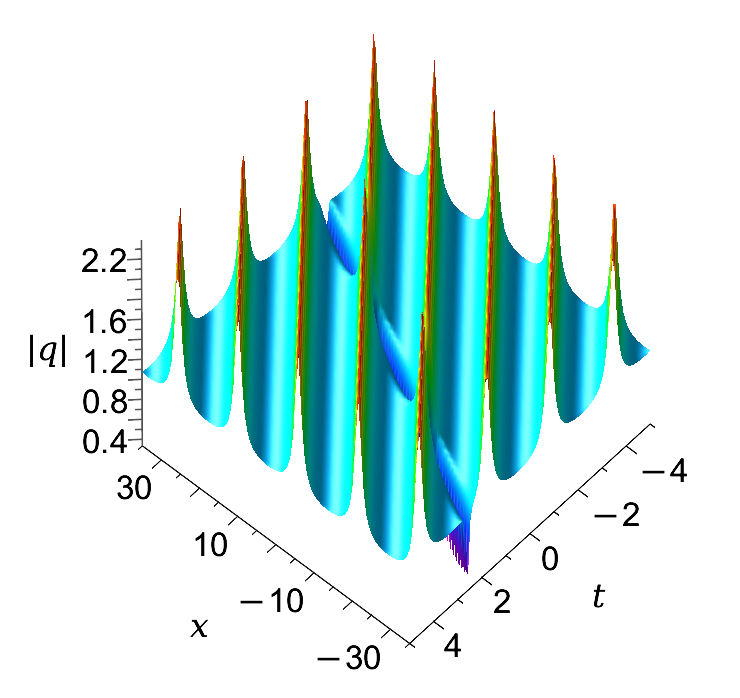}
        \caption{}
        \label{fg63}
    \end{subfigure}
     \caption{The interaction solutions with $a=c=c_1=c_2=1$. (a) The double-periodic wave with $\rho=1,b=-\frac{15}{2},\lambda_1=4\mathrm{i},\lambda_2=3\mathrm{i}$; (b) The periodic wave and the bright soliton with $\rho=0,b=\frac{11}{2},\lambda_1=-\mathrm{i},\lambda_2=3\mathrm{i}$; (c) The periodic wave and the dark soliton with $\rho=0,b=\frac{11}{2},\lambda_1=\mathrm{i},\lambda_2=3\mathrm{i}$.}
    \label{fg6}
\end{figure}

Secondly, a double-periodic wave solution is derived in Fig.\ref{fg61}. Notably, double-periodic wave solutions can also be obtained when $\mathrm{Im}(\lambda_1)=\mathrm{Im}(\lambda_2)=0$. Thirdly, as presented in Fig.\ref{fg62} and Fig.\ref{fg63}, the interactions between periodic waves and bright or dark solitons are found using the single-fold DT.

\begin{figure}[htbp]
    \centering
   \begin{subfigure}{0.31\textwidth}
        \centering
        \includegraphics[width=\textwidth]{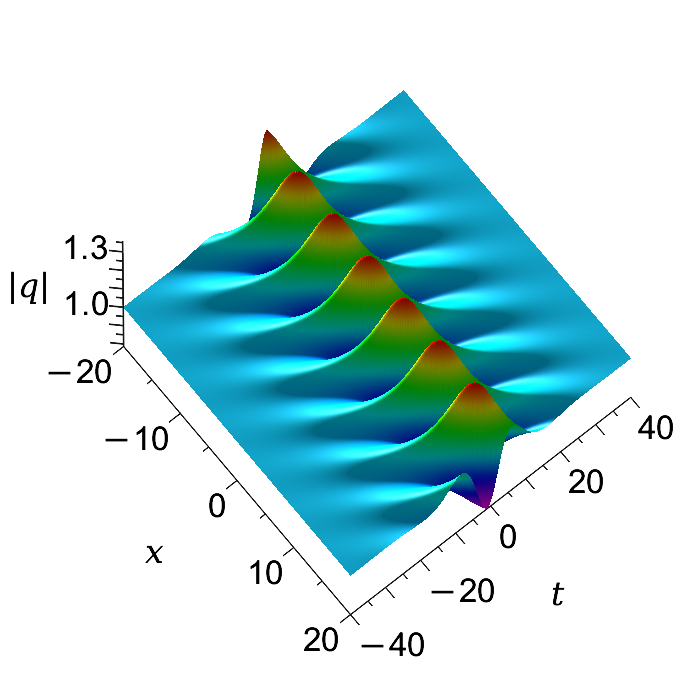}
        \caption{}
        \label{fg71}
    \end{subfigure}
  %  \hspace{1cm}
    \begin{subfigure}{0.32\textwidth}
        \centering
        \includegraphics[width=\textwidth]{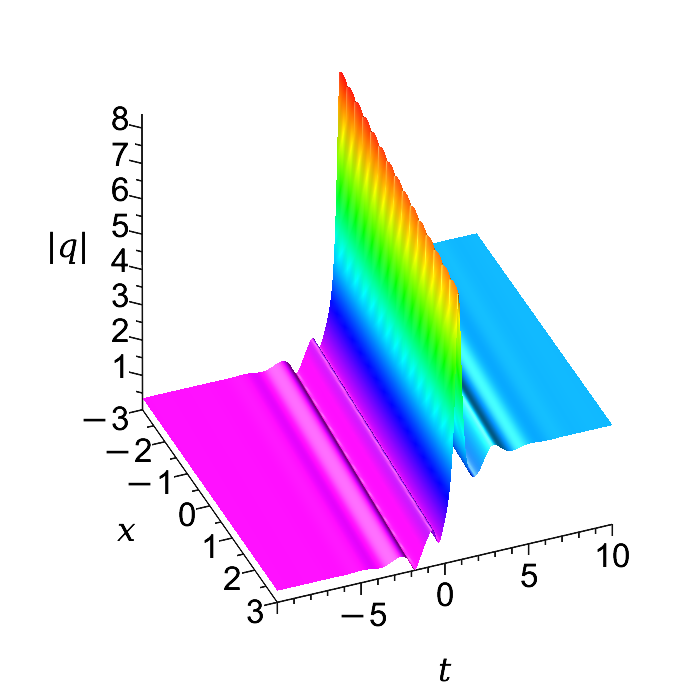}
        \caption{}
        \label{fg72}
    \end{subfigure}
    \begin{subfigure}{0.35\textwidth}
        \centering
        \includegraphics[width=\textwidth]{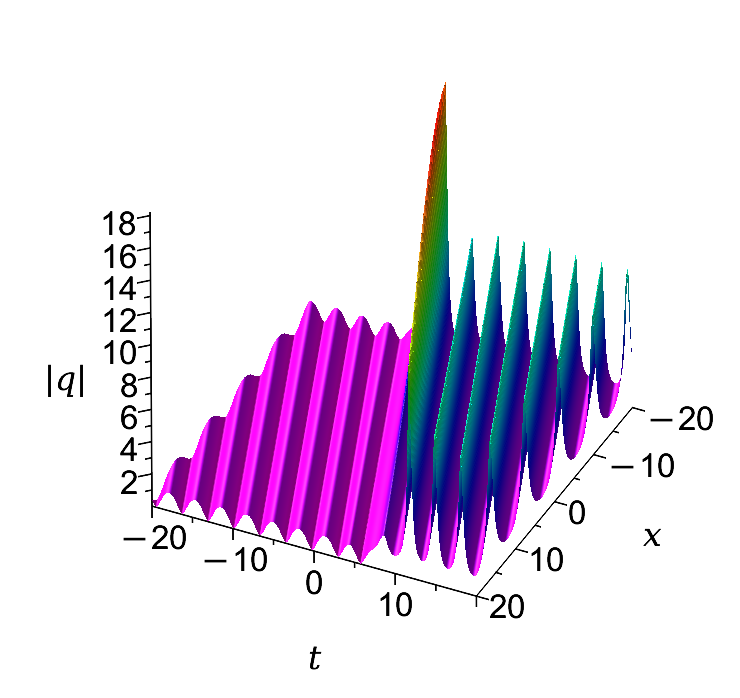}
        \caption{}
        \label{fg73}
    \end{subfigure}
     \caption{(a) The breather solution with $\rho=1,a=1,b=-\frac{15}{2},\lambda_1=\lambda_2^*=2+5\mathrm{i}$; (b) Mixed kink and soliton solution with $\rho=-1,a=-1,b=-\frac{21}{2},\lambda_1=2\mathrm{i},\lambda_2=\frac{1}{2}-2\mathrm{i}$; (c) Mixed kink and periodic wave solution with $\rho=-1,a=-1,b=-\frac{21}{2},\lambda_1=4\mathrm{i},\lambda_2=\frac{1}{3}-3\mathrm{i}$.}
    \label{fg7}
\end{figure}

Furthermore, when $\lambda_1=\lambda_2^*$, a breather solution is obtained in Fig.\ref{fg71}. Under the condition $\mathrm{Re}(\lambda_1)=0,\mathrm{Re}(\lambda_2)\neq0$, there exist mixed solutions among the kink, the soliton and periodic wave as depicted in Fig.\ref{fg72} and Fig.\ref{fg73}. As shown, the soliton component with the largest amplitude, embedded in the kink solution, decays exponentially over time.

In order to derive rogue wave and related interaction solutions, we introduce $\lambda_j=\lambda_j+\epsilon^2$ and the above eigenfunctions \eqref{eg} can be expanded in a Taylor series as follows,
\begin{align*}
   & \lambda_j^{m}\phi_{k,j}=\phi_{k,j}^{[m]}[0]+\phi_{k,j}^{[m]}[1]\epsilon^2+\phi_{k,j}^{[m]}[2]\epsilon^4+\cdots+\phi_{k,j}^{[m]}[n]\epsilon^{2n}+\cdots, \\
    &\phi_{k,j}^{[m]}[n]=\frac{1}{(2n)!}\frac{\partial^{2n}}{\partial\epsilon^{2n}}[ (\lambda_j+\epsilon^2)^{m}\phi_{k,j}(\lambda_j+\epsilon^2)]_{\epsilon=0},~k=1,2.
\end{align*}
Similar to the construction method of degenerate DT as described in \cite{dege}, let $\lambda_{2j-1}=\lambda_{2j}^*=\frac{\sqrt{2(-c^2+\sqrt{c^4+2ac^2-4c^2\rho}-a+2\rho)}}{a-2\rho},j=1,2,3,\cdots,N$, when $c_1=\frac{1}{\epsilon}, c_2=-\frac{1}{\epsilon}$, the $N$-order rogue wave solution is derived as follows,
\begin{align}\label{gb1}
    q[N]=q-2i\rho A^{'}_{2,N-1}+A^{'}_{2,N-1,x},
\end{align}
where $A^{'}_{2,N-1}=\frac{|B^{'}_{11}|}{|B^{'}_{12}|}$ with $B^{'}_{12}=$
\[ 
    \begin{pmatrix}
        \phi_{1,1}^{[2N-2]}[0] & \phi_{2,1}^{[2N-1]}[0] &  \phi_{1,1}^{[2N-4]}[0] &
        \cdots & \phi_{1,1}^{[0]}[0] & \phi_{2,1}^{[1]}[0] \\
         \phi_{1,2}^{[2N-2]}[0] & \phi_{2,2}^{[2N-1]}[0] &  \phi_{1,2}^{[2N-4]}[0] &
     \cdots & \phi_{1,2}^{[0]}[0] & \phi_{2,2}^{[1]}[0]  \\
        \vdots & \vdots & \vdots & \ddots & \vdots & \vdots \\
        \phi_{1,2N-1}^{[2N-2]}[N-1] &  \phi_{2,2N-1}^{[2N-1]}[N-1] &  \phi_{1,2N-1}^{[2N-4]}[N-1] & \cdots & \phi_{1,2N-1}^{[0]}[N-1] & \phi_{2,2N-1}^{[1]}[N-1] \\
        \phi_{1,2N}^{[2N-2]}[N-1] &  \phi_{2,2N}^{[2N-1]}[N-1] &  \phi_{1,2N}^{[2N-4]}[N-1] & \cdots & \phi_{1,2N}^{[0]}[N-1] & \phi_{2,2N}^{[1]}[N-1] 
    \end{pmatrix},\]
and $B^{'}_{11}$ is obtained by replacing the second column of $B^{'}_{12}$ with 
$$(-\phi_{1,1}^{[2N]}[0], -\phi_{1,2}^{[2N]}[0], ,\cdots,-\phi_{1,2N-1}^{[2N]}[N-1],-\phi_{1,2N}^{[2N]}[N-1] )^{T}.$$
Correspondingly, when $\lambda_{2j-1}=\lambda_{2j}^*=\frac{\sqrt{2(-c^2+\sqrt{c^4+2ac^2-4c^2\rho}-a+2\rho)}}{a-2\rho},j=1,2,3,\cdots,n$, the interaction solutions between an $n$-order $(n<N)$ rogue wave and other localized waves are obtained as follows, 
\begin{align}\label{gb2}
    q[N]=q-2i\rho A^{''}_{2,N-1}+A^{''}_{2,N-1,x},
\end{align}
where $A^{''}_{2,N-1}=\frac{|B^{''}_{11}|}{|B^{''}_{12}|}$ with $B^{''}_{12}=$
\[ 
    \begin{pmatrix}
        \phi_{1,1}^{[2N-2]}[0] & \phi_{2,1}^{[2N-1]}[0] &  \phi_{1,1}^{[2N-4]}[0] &
        \cdots & \phi_{1,1}^{[0]}[0] & \phi_{2,1}^{[1]}[0] \\
         \phi_{1,2}^{[2N-2]}[0] & \phi_{2,2}^{[2N-1]}[0] &  \phi_{1,2}^{[2N-4]}[0] &
     \cdots & \phi_{1,2}^{[0]}[0] & \phi_{2,2}^{[1]}[0]  \\
        \vdots & \vdots & \vdots & \ddots & \vdots & \vdots \\
        \phi_{1,2n-1}^{[2N-2]}[n-1] &  \phi_{2,2n-1}^{[2N-1]}[n-1] &  \phi_{1,2n-1}^{[2N-4]}[n-1] & \cdots & \phi_{1,2n-1}^{[0]}[n-1] & \phi_{2,2n-1}^{[1]}[n-1] \\
        \phi_{1,2n}^{[2N-2]}[n-1] &  \phi_{2,2n}^{[2N-1]}[n-1] &  \phi_{1,2n}^{[2N-4]}[n-1] & \cdots & \phi_{1,2n}^{[0]}[n-1] & \phi_{2,2n}^{[1]}[n-1] \\
        \lambda_{2n+1}^{2N-2}\phi_{1,2n+1} &  \lambda_{2n+1}^{2N-1}\phi_{2,2n+1} &
         \lambda_{2n+1}^{2N-4}\phi_{1,2n+1} & \cdots &  \phi_{1,2n+1} &  \lambda_{2n+1}\phi_{2,2n+1} \\
         \lambda_{2n+2}^{2N-2}\phi_{1,2n+2} &  \lambda_{2n+2}^{2N-1}\phi_{2,2n+2} &
         \lambda_{2n+2}^{2N-4}\phi_{1,2n+2} & \cdots &  \phi_{1,2n+2} &  \lambda_{2n+2}\phi_{2,2n+2}\\
          \vdots & \vdots & \vdots & \vdots & \vdots & \vdots \\
          \lambda_{2N}^{2N-2}\phi_{1,2N} &  \lambda_{2N}^{2N-1}\phi_{2,2N} &
         \lambda_{2N}^{2N-4}\phi_{1,2N} & \cdots &  \phi_{1,2N} &  \lambda_{2N}\phi_{2,2N}
    \end{pmatrix},\]
and $B^{''}_{11}$ is obtained by replacing the second column of $B^{''}_{12}$ with 
$$(-\phi_{1,1}^{[2N]}[0], -\phi_{1,2}^{[2N]}[0], ,\cdots,-\phi_{1,2n}^{[2N]}[n-1],\lambda_{2n+1}^{2N}\phi_{1,2n+1}, \cdots,\lambda_{2N}^{2N}\phi_{1,2N})^{T}.$$

\begin{figure}[htbp]
    \centering
   \begin{subfigure}{0.25\textwidth}
        \centering
        \includegraphics[width=\textwidth]{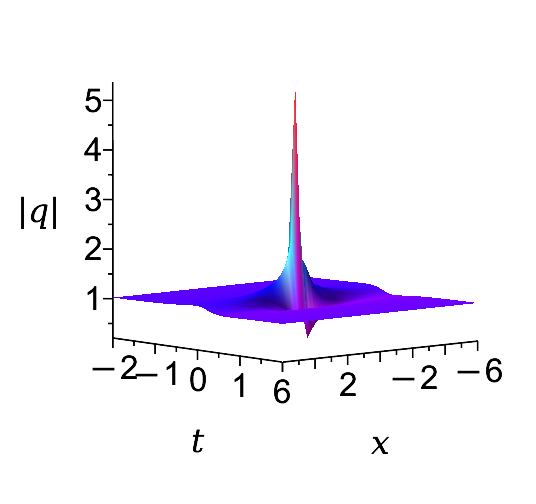}
        \caption{}
        \label{fg81}
    \end{subfigure}
  %  \hspace{1cm}
    \begin{subfigure}{0.24\textwidth}
        \centering
        \includegraphics[width=\textwidth]{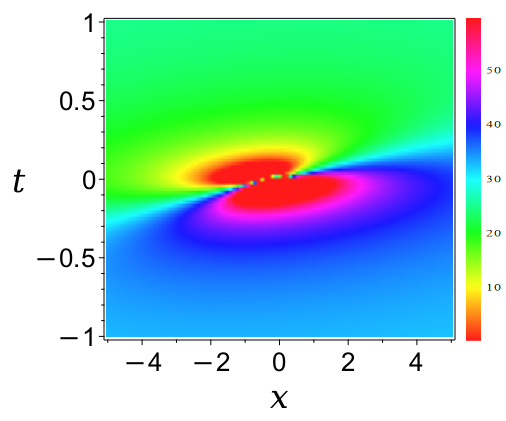}
        \caption{}
        \label{fg82}
    \end{subfigure}
    \begin{subfigure}{0.24\textwidth}
        \centering
        \includegraphics[width=\textwidth]{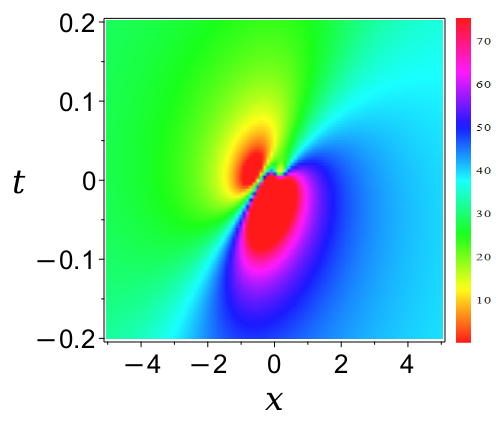}
        \caption{}
        \label{fg83}
    \end{subfigure}
        \begin{subfigure}{0.24\textwidth}
        \centering
        \includegraphics[width=\textwidth]{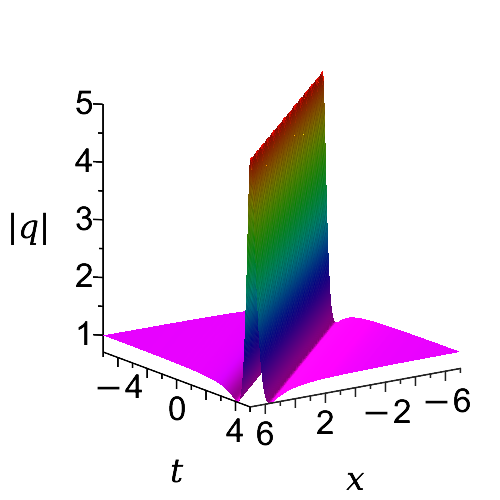}
        \caption{}
        \label{fg88}
    \end{subfigure}
      \caption{(a) The first-order rogue wave with $\rho = 1, a = -2, c = 1, b = 18,\lambda_1=\lambda_2^*=\frac{\sqrt{7}}{4}-\frac{1}{4}\mathrm{i}$; (b)(c) The density plots of (a); (d) The line rogue wave with $\rho = 0, a = -1, c = 1, b =\frac{1}{2},\lambda_1=\lambda_2^*=1-\mathrm{i}$.}
    \label{fg8}
 \end{figure}

Based on the expression \eqref{gb1}, when $N=1$, the first-order rogue wave solution is derived in Fig.\ref{fg81} and Fig.\ref{fg82} by choosing the appropriate parameters. Its exact expression is 
\begin{equation*}
q_r=\frac{2e^{2\mathrm{i}(9t-x)}Q}{(390\mathrm{i}t\sqrt{7} + 4\mathrm{i}\sqrt{7}\,x + \sqrt{7} + 11\mathrm{i} - 770t + 84x)^2},
\end{equation*}
where $Q=300300\mathrm{i}\sqrt{7}\,t^2 - 29680\mathrm{i}\sqrt{7}\,t x - 336\mathrm{i}\sqrt{7}\,x^2 + 125\mathrm{i}\sqrt{7} + 5740\sqrt{7}\,t - 952\sqrt{7}\,x- 35420\mathrm{i}\,t + 280\mathrm{i}\,x + 235900t^2 + 75600t x - 3472x^2 + 97$.
It is obvious to find that the first-order rogue wave has one peak and one valley, and moreover, the amplitudes of the plane wave backgrounds for the wave peak and the wave valley are different. Additionally, a fundamental line rogue wave (W-shaped rational soliton) is obtained in Fig.\ref{fg88}, with the exact expression
\begin{align*}
  q_{l}=\frac{[18\mathrm{it^2+(-24\mathrm{i}x-12-12\mathrm{i})t+8\mathrm{i}x^2+8(1+\mathrm{i})x+4-8\mathrm{i}}]e^{\frac{1}{2}\mathrm{i}(t-2x)}}{(-3t+2x+2+3\mathrm{i}t-2\mathrm{i}x)^2}.
\end{align*}
 
\begin{figure}[ht!]
     \begin{subfigure}{0.31\textwidth}
        \centering
        \includegraphics[width=\textwidth]{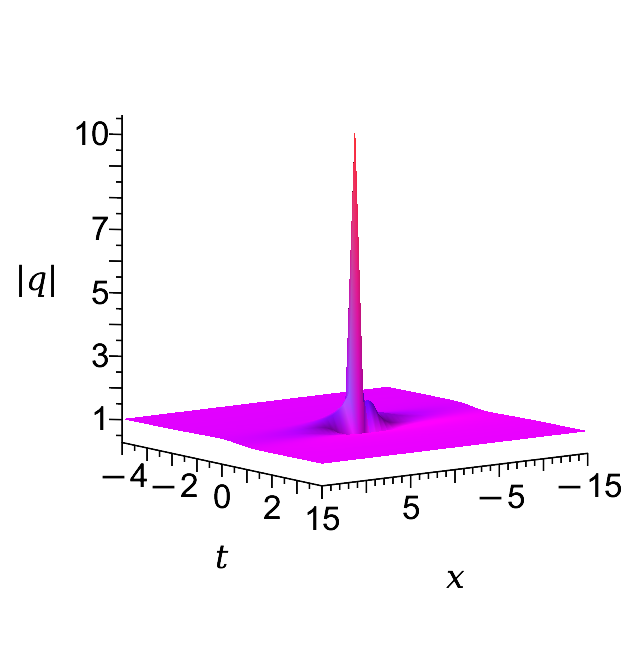}
        \caption{}
        \label{fg84}
    \end{subfigure}
        \begin{subfigure}{0.31\textwidth}
        \centering
        \includegraphics[width=\textwidth]{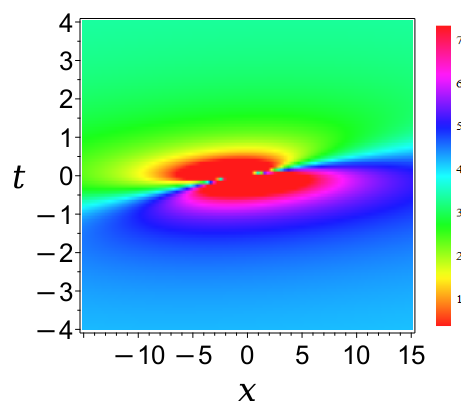}
        \caption{}
        \label{fg85}
    \end{subfigure}
        \begin{subfigure}{0.33\textwidth}
        \centering
        \includegraphics[width=\textwidth]{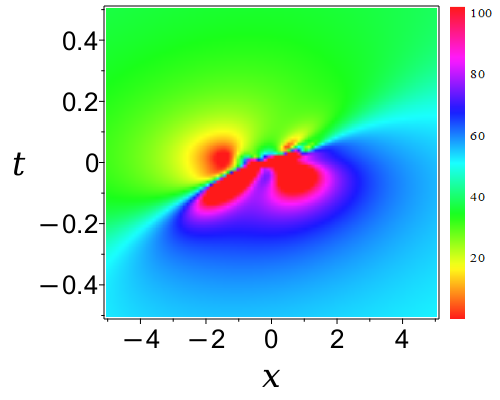}
        \caption{}
        \label{fg86}
    \end{subfigure}
     \caption{(a) The second-order rogue wave with $\rho = 1, a = -2, c = 1, b = 18,\lambda_1=\lambda_3=\lambda_2^*=\lambda_4^*=\frac{\sqrt{7}}{4}-\frac{1}{4}\mathrm{i}$; (b)(c) The density plots of (a).}
    \label{fg9}
\end{figure}

Consequently, for $N=2$, the second-order rogue wave is obtained in Fig.\ref{fg84} and Fig.\ref{fg85} from the formula \eqref{gb1}. When viewed on a large scale, its profile resembles the first-order rogue wave. However, a small-scale view reveals distinct differences. This asymmetric waveform is a key feature distinguishing it from local equations, and its amplitude is approximately twice that of the first-order rogue wave. 

\begin{figure}[ht!]
     \begin{subfigure}{0.31\textwidth}
        \centering
        \includegraphics[width=\textwidth]{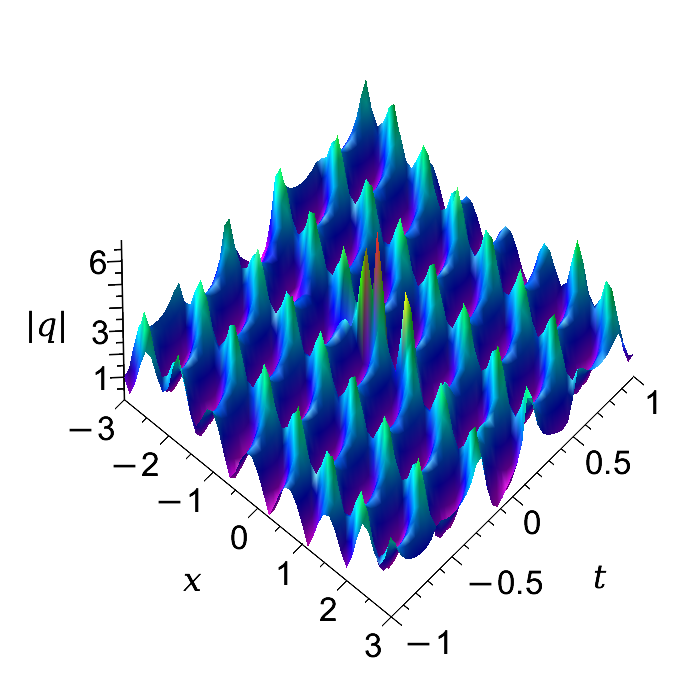}
        \caption{}
        \label{fg101}
    \end{subfigure}
        \begin{subfigure}{0.31\textwidth}
        \centering
        \includegraphics[width=\textwidth]{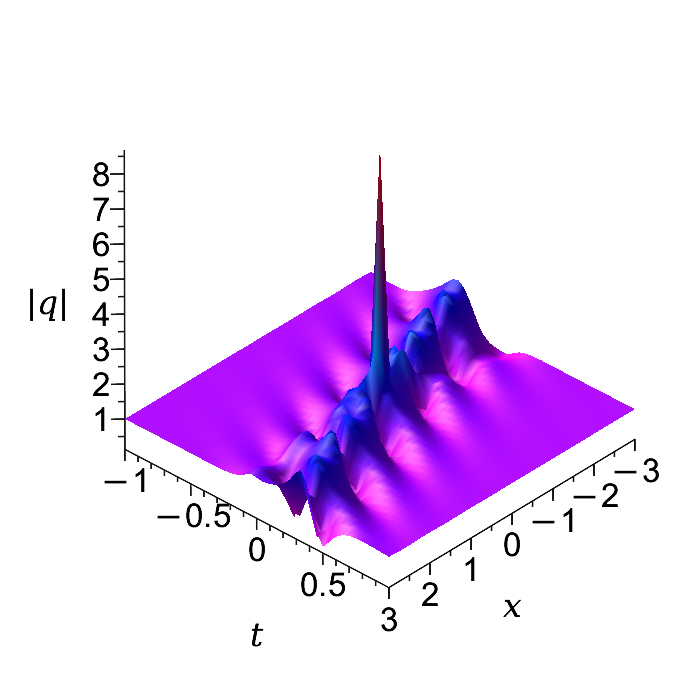}
        \caption{}
        \label{fg102}
    \end{subfigure}
        \begin{subfigure}{0.31\textwidth}
        \centering
        \includegraphics[width=\textwidth]{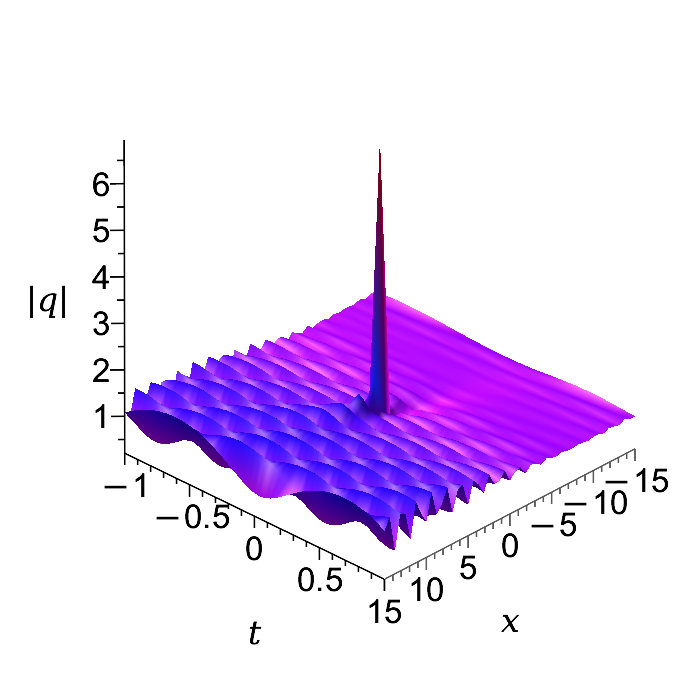}
        \caption{}
        \label{fg103}
    \end{subfigure}
     \caption{The interaction solutions with $\rho = 1, a = -4, c = 1, b = -5$. (a) The double-periodic wave and rogue wave with $\lambda_1=\lambda_2^*=\frac{\sqrt{11}}{6}-\frac{1}{6}\mathrm{i},\lambda_3=1,\lambda_4=2$; (b) The breather and rogue wave with $\lambda_1=\lambda_2^*=\frac{\sqrt{11}}{6}-\frac{1}{6}\mathrm{i},\lambda_3=\lambda_4^*=1-\mathrm{i}$; (c) The kink, periodic wave and rogue wave with $\lambda_1=\lambda_2^*=\frac{\sqrt{11}}{6}-\frac{1}{6}\mathrm{i},\lambda_3=2\mathrm{i},\lambda_4=\frac{1}{2}-2\mathrm{i}$.}
    \label{fg10}
\end{figure}

Based on the expression \eqref{gb2} with $N=2,n=1$, interactions between a first-order rogue wave and other localized wave are obtained. In Fig.\ref{fg101}, we present the interaction between a first-order rogue wave and a double-periodic wave. The interaction between a first-order rogue wave and a breather is derived in Fig.\ref{fg102}. Fig. \ref{fg103} shows a composite solution of a rogue wave, a kink, and a periodic wave. Similarly, more types of interaction solutions can be derived through the selection of appropriate parameters.

\section{Conclusions and discussions}\label{4s}
In conclusion, we have systematically investigated the reverse space-time higher-order mNLS equation. The following key conclusions are drawn:

The Lax pair and infinitely many conservation laws have been established, confirming the complete integrability of this nonlocal system. By successfully constructing the Darboux transformation, we have derived a wide variety of mixed soliton solutions and localized wave structures. 

Under the nonlocal symmetry, some dynamical behaviors have no counterpart in the local equation. For instance, kink solutions, exponentially decaying solitons, asymmetric rogue waves and their interactions exist in nonlocal equations but are not observed in local equations. In addition, compared to the local case, the $N$-fold DT of the nonlocal equation incorporates $2N$ distinct spectral parameters, leading to significantly richer solution structures.

In summary, this work not only provides a systematic solution scheme for the studied equation, but also offers new insights into wave dynamics in parity-time symmetric or other nonlocal systems.

%\section*{Acknowledgments}
%The authors thanks Prof. Q. P. Liu for helpful and enlightening suggestions. 

\section*{Data availability}
Data sharing is not applicable to this article as no new data were
created or analyzed in this study.

\section*{Funding}
The authors have not disclosed any funding.
\section*{Author Declarations}
The author declare that there is no conflict of interest regarding the publication of this paper.

% 进入附录后重新定义公式编号格式
\section*{Appendix}\label{app:A}
\renewcommand{\theequation}{A.\arabic{equation}}
\setcounter{equation}{0} % 重置公式计数器
Ref.\cite{steep} displays a new Lax integrable hierarchy as follows,
\begin{align*}
    q_t+2\mathrm{i}\rho M_2^{(-1)}-M_{2,x}^{(-1)}-2qM_1^{(0)}-2\mathrm{i}M_2^{(0)}=0,\\
    r_t-2\mathrm{i}\rho M_3^{(-1)}-M_{3,x}^{(-1)}+2rM_1^{(0)}+2\mathrm{i}M_3^{(0)}=0,
\end{align*}
where \begin{align*}
    M_2^{(-m)}&=\mathrm{i}\alpha q, ~~M_3^{(-m)}=\mathrm{i}\alpha r,~~M_1^{(-m+1)}=\frac{1}{2}\alpha qr+\beta,\\
    M_2^{(-m+1)}&=\mathrm{i}\alpha\rho q-\frac{\alpha q_x}{2}+\frac{1}{2}\mathrm{i}q^2r+\mathrm{i}\beta q,~~M_3^{(-m+1)}=\mathrm{i}\alpha\rho r+\frac{\alpha r_x}{2}+\frac{1}{2}\mathrm{i}qr^2+\mathrm{i}\beta r,\\
    M_1^{(-m+2)}&=(\alpha\rho+\frac{1}{2}\beta)qr+\frac{1}{4}\mathrm{i}\alpha(q_xr-qr_x)+\frac{3}{8}\alpha q^2r^2+\xi,\\
    M_2^{(-m+2)}&=\alpha(\frac{3}{8}\mathrm{i}q^3r^2-\frac{3}{4}qq_xr+\frac{3}{2}\mathrm{i}\rho q^2r-\rho q_x+\mathrm{i}\rho^2q-\frac{1}{4}\mathrm{i}q_{xx})\\
    &\quad+\beta(\frac{1}{2}\mathrm{i}q^2r-\frac{1}{2}q_x+\mathrm{i}\rho q)+\mathrm{i}\xi q,~~M_3^{(-m+2)}=\beta(\frac{1}{2}\mathrm{i}qr^2+\frac{1}{2}r_x+\mathrm{i}\rho r)\\&\quad\alpha(\frac{3}{8}\mathrm{i}q^2r^3+\frac{3}{4}qr_xr+\frac{3}{2}\mathrm{i}\rho qr^2+\rho r_x+\mathrm{i}\rho^2r-\frac{1}{4}\mathrm{i}r_{xx})+\mathrm{i}\xi r,\\ M_1^{(m)}&=M_2^{(m)}=M_3^{(m)}=M_2^{(0)}=M_3^{(0)}=0,~~M_1^{(0)}=\gamma, 
\end{align*}
and $\alpha,\beta,\gamma,\xi$ are the arbitrary constants. 

When we choose $\alpha=-4\mathrm{i},\beta=8\mathrm{i}\rho,\gamma=-4\mathrm{i}\rho^2,\xi=0$, the above hierarchy is reduced to an integrable coupled system,
\begin{align}\label{eq:app1}
    \mathrm{i}q_t&+\mathrm{i}q_{xxx}+(3qr+2\rho)q_{xx}+3rq_{x}^2+\frac{1}{2}(-9\mathrm{i}q^2r^2-20\mathrm{i}\rho qr+8\mathrm{i}\rho^2+6qr_x)q_x \notag \\&+\mathrm{i}(-3q^3r-2\rho q^2)r_x-3\rho r^2q^3-4\rho^2rq^2+8(\rho^3-\rho^2)q,\notag\\
    \mathrm{i}r_t&+\mathrm{i}r_{xxx}-(3qr+2\rho)r_{xx}-3qr_{x}^2+\frac{1}{2}(-9\mathrm{i}q^2r^2-20\mathrm{i}\rho qr+8\mathrm{i}\rho^2-6rq_x)r_x \notag \\&+\mathrm{i}(-3r^3q-2\rho r^2)q_x+3\rho r^3q^2+4\rho^2r^2q+8(-\rho^3+\rho^2)r.
  \end{align}
The corresponding Lax pair for Eq.\eqref{eq:app1} is given by
\begin{align}\label{eq:app2}
\Psi_x=
P\Psi, ~~~\Psi_t=Q
\Psi,
\end{align}
where 
\begin{align*}
P&=\begin{pmatrix}
-\dfrac{i}{\lambda^2} + \rho i & \dfrac{q}{\lambda} \\
\dfrac{r}{\lambda} & \dfrac{i}{\lambda^2} - \rho i
\end{pmatrix}, ~~Q=\begin{pmatrix}
Q_1 & Q_2 \\
Q_3 & -Q_1
\end{pmatrix},\\
Q_1&=-\frac{3i q^2 r^2}{2}\lambda^{-2} - 4i \rho^2 - r_x \, q\lambda^{-2} + q_x \, r\lambda^{-2} - 2i q r\lambda^{-4} + 8i \rho\lambda^{-4} - 4i\lambda^{-6}, \\ Q_2&=-q_{xx}\lambda^{-1} + 3i q_x q r\lambda^{-1} + 2i q_x\lambda^{-3} + \frac{3 q^3 r^2}{2}\lambda^{-1} +2 q^2 r \rho\lambda^{-1} + 2 q^2 r\lambda^{-3}\\&\quad - 4 q \rho^2\lambda^{-1} - 4 q \rho\lambda^{-3} + 4 q\lambda^{-5}, \\Q_3&=-r_{xx}\lambda^{-1} -3ir_x q r\lambda^{-1} - 2i r_x\lambda^{-3} + \frac{3 q^2 r^3}{2}\lambda^{-1} + 2 q r^2 \rho\lambda^{-1} + 2 q r^2\lambda^{-3}\\&\quad - 4 r \rho^2\lambda^{-1} - 4 r \rho\lambda^{-3} + 4 r\lambda^{-5}.
\end{align*}

\end{document}